\documentclass[twocolumn,amsmath,aps,fleqn]{revtex4}

\usepackage{graphicx}

\usepackage{hyperref} 

\newcommand{\adsurl}[1]{\href{#1}{ADS}} 
\providecommand{\url}[1]{\href{#1}{#1}}

\makeatletter
    \newcommand\figcaption{\def\@captype{figure}\caption}
    \newcommand\tabcaption{\def\@captype{table}\caption}
\makeatother
\newcommand{\bd}{{\bf d}}
\newcommand{\rF}{{\rm F}}

\newcommand\eq[1]{Eq.~(\ref{#1})}

\newcommand{\be}{\begin{equation}}
\newcommand{\ee}{\end{equation}}
\newcommand{\bea}{\begin{eqnarray}}
\newcommand{\eea}{\end{eqnarray}}
\newcommand{\rv}{{\rm v}}
\newcommand{\bk}{{\bf k}}
\newcommand{\rd}{{\rm d}}

\newcommand{\bv}{{\bf v}}

\newcommand{\br}{{\bf r}}
\newcommand{\bri}{{\bf r_{\bf i}}}
\newcommand{\brj}{{\bf r_{\bf j}}}
\newcommand{\brij}{{\bf r_{\bf ij}}}

\newcommand{\hbri}{{\hat{\bf r}_{\bf i}}}
\newcommand{\hbrj}{{\hat{\bf r}_{\bf j}}}
\newcommand{\hbrij}{{\hat{\bf r}_{\bf ij}}}
\newcommand{\hbr}{\hat{\bf r}}
\renewcommand{\(}{\left(}
\renewcommand{\)}{\right)}
\newcommand{\<}{\left<}
\renewcommand{\>}{\right>}
\renewcommand{\d}{\rm d}

\begin{document}


\title{Cosmological Constraints from Type Ia Supernovae Peculiar Velocity Measurements}
\author{C. Gordon}
\affiliation{Oxford Astrophysics, Physics, DWB, Keble Road, Oxford, OX1 3RH, UK}

\author{K. Land}
\affiliation{Oxford Astrophysics, Physics, DWB, Keble Road, Oxford, OX1 3RH, UK}

\author{A. Slosar}
\affiliation{Oxford Astrophysics, Physics, DWB, Keble Road, Oxford, OX1 3RH, UK}

\begin{abstract}
We detect the correlated peculiar velocities of nearby type Ia supernovae (SNe), 
while highlighting an error in some of the literature.  We find 
$\sigma_8=0.79 \pm 0.22$ from SNe, and examine the potential of this method to constrain 
cosmological parameters in the future. We demonstrate that a survey of 300 low-$z$ SNe 
(such as the nearby SNfactory) will underestimate the errors on $w$ by $\sim$35\% if 
the coherent peculiar velocities are not included.
\end{abstract}

\maketitle




The  first compelling evidence that the Universe is undergoing a period of 
accelerated expansion was provided by observations of Type Ia supernovae (SNe)
~\citep{Riess:1998cb,Perlmutter:1998np}.
The data from many current~\citep{Astier:2005qq,Riess:2006fw,essence}
and near future~\citep{SNfact,sdsssn,carnegysn,SNAP2} surveys
should eventually constrain the effective dark energy equation of state to 
better than 10\%.

Density inhomogeneities cause the SNe to deviate from the Hubble flow, as 
gravitational instability leads to matter flowing out of under-densities and 
into over-densities. These ``peculiar velocities'' (PVs) lead to an increased scatter 
in the Hubble diagram, of which several studies have been made~\citep{rieprekir95,riess97,
zehriekir98,bonacic00,radluchud04,bonvin,haugboelle06,
jhariekir06,watfeld07,conley07,wang07,nehuco07,Han}. 
When combining low and high redshift SNe in order to estimate the properties of 
the dark energy, the velocity contributions are usually modeled as a Gaussian 
noise term which is uncorrelated between different SNe. However, as recently 
emphasized~\citep{Hui:2005nm}, in the limit of low redshift $z\lesssim 0.1$ and 
large sample size, the correlations between SNe PVs contribute significantly 
to the overall error budget.
In this letter we investigate the effect of incorporating these correlations 
using the largest available low redshift compilation~\citep{jhariekir06}. 



\section{Peculiar velocity covariance}\label{formula}
\vspace{-0.4cm}
The luminosity distance, $d_L$, to a SN at redshift $z$, is defined such that
${\mathcal F} = \frac{\mathcal L}{4\pi d_L^2}$
where $\mathcal F$ is the observed flux and $\mathcal L$ is the SN's intrinsic
luminosity. Astronomers use magnitudes, which are related to 
the luminosity distance (in megaparsec) by
\be
\mu \equiv m - M = 5\log_{10} d_L + 25\label{dlobs},
\ee
where $m$ and $M$ are the apparent and absolute magnitudes respectively.
In the context of SNe, $M$ is a ``nuisance parameter'' which is degenerate 
with $\log(H_0)$ and can be marginalised over. 
For a Friedmann-Robertson-Walker Universe
the predicted luminosity distance is given by
\be
d_{L}(z)= (1+z)\int_{0}^z \frac{{\rd}z'}{H(z')}\label{dlth}
\ee 
in speed of light units, where $H$ is the Hubble parameter.
In the limit of low redshift this reduces to $d_{L}\approx z/H_{0}$.

Assuming a flat Universe (as we shall do throughout this letter), the effect of peculiar velocities leads to 
a perturbation in the luminosity distance ($\delta d_{L}$)  given by
~\citep{sasaki87,sugsugsas99,pynbirk95,bondurgas05,Hui:2005nm}
\begin{equation}
\frac{\delta d_{L}}{d_{L}} = \hbr \cdot \(\bv-\frac{(1+z)^{2}}{ 
H(z) \: d_{L}}[\bv-\bv_O]\)
\label{ddod}
\end{equation}
where $\br$ is the position of the SN, 
and $\bv_O$ and $\bv$ are the peculiar velocites of the observer and SN repectively. 
Using the Cosmic Microwave Background dipole we can very accurately correct 
for $\bv_O$. 
This demonstrates how a SNe survey that measures $\mu$ and $z$ can 
estimate the projected PV field. We now relate this to the cosmology.

The projected velocity correlation function, 
$\xi(\bri,\brj)\equiv \< (\bv(\bri)\cdot\hbri)( \bv(\brj)\cdot\hbrj)\>$, 
must be rotationally invariant, and therefore it can be 
decomposed into a parallel and perpendicular 
components~\citep{Gorski,Groth,Scott} :
\be
\xi(\bri,\brj)=\sin\theta_{i}\sin\theta_j\xi_{\perp}(r,z_i,z_j)+\cos\theta_{i}
\cos\theta_{j}\xi_{\parallel}(r,z_i,z_j)
\notag
\ee
where $\brij \equiv \bri-\brj$, $r=|\brij|$, $\cos\theta_{i}\equiv\hbri\cdot\hbrij$, 
and $\cos\theta_{j}\equiv\hbrj\cdot\hbrij$. In linear theory, 
these are given by~\citep{Gorski,Groth,Scott}:
\be
\xi_{\parallel,\perp}
= D'(z_i)\:D'(z_j)\: \int_{0}^{\infty} \frac{\d k}{2\pi^{2}}P(k) 
K_{\parallel,\perp}(k r)\label{xith}
\ee
where  for an arbitrary variable $x$, 
$K_{\parallel}(x)\equiv j_0(x)-\frac{2j_1(x)}{x}$, $K_{\perp}(x)\equiv j_1(x)/x$. 
$D(z)$ is the growth function, and derivatives are with respect to conformal 
time. $P(k)$ is the matter power spectrum. This corrects the formulae used in
~\citep{Monteagudo:2005ys,Cooray:2006ft}, see the Appendix for details.

The above estimate of $\xi(\bri,\brj)$ is based on linear theory.
On scales smaller than about 10$h^{-1}$Mpc nonlinear contributions become important.
These are usually 
modeled as an uncorrelated term which is independent of redshift, 
often set to $\sigma_{v}\sim 300$ km/s. 
Comparison with numerical simulations \citep{silberman01} has confirmed that this is an
 effective way of accounting for the nonlinearities.
Other random errors that are usually considered 
are those from the lightcurve fitting ($\mu_{\rm err}$), 
and  intrinsic magnitude scatter ($\sigma_m$) found to be 
$\sim 0.08$ in the case of~\cite{jhariekir06}. It is just these three errors that are usually included in the 
analysis of SNe.

\begin{figure}
\centerline{\includegraphics[width=8.0cm]{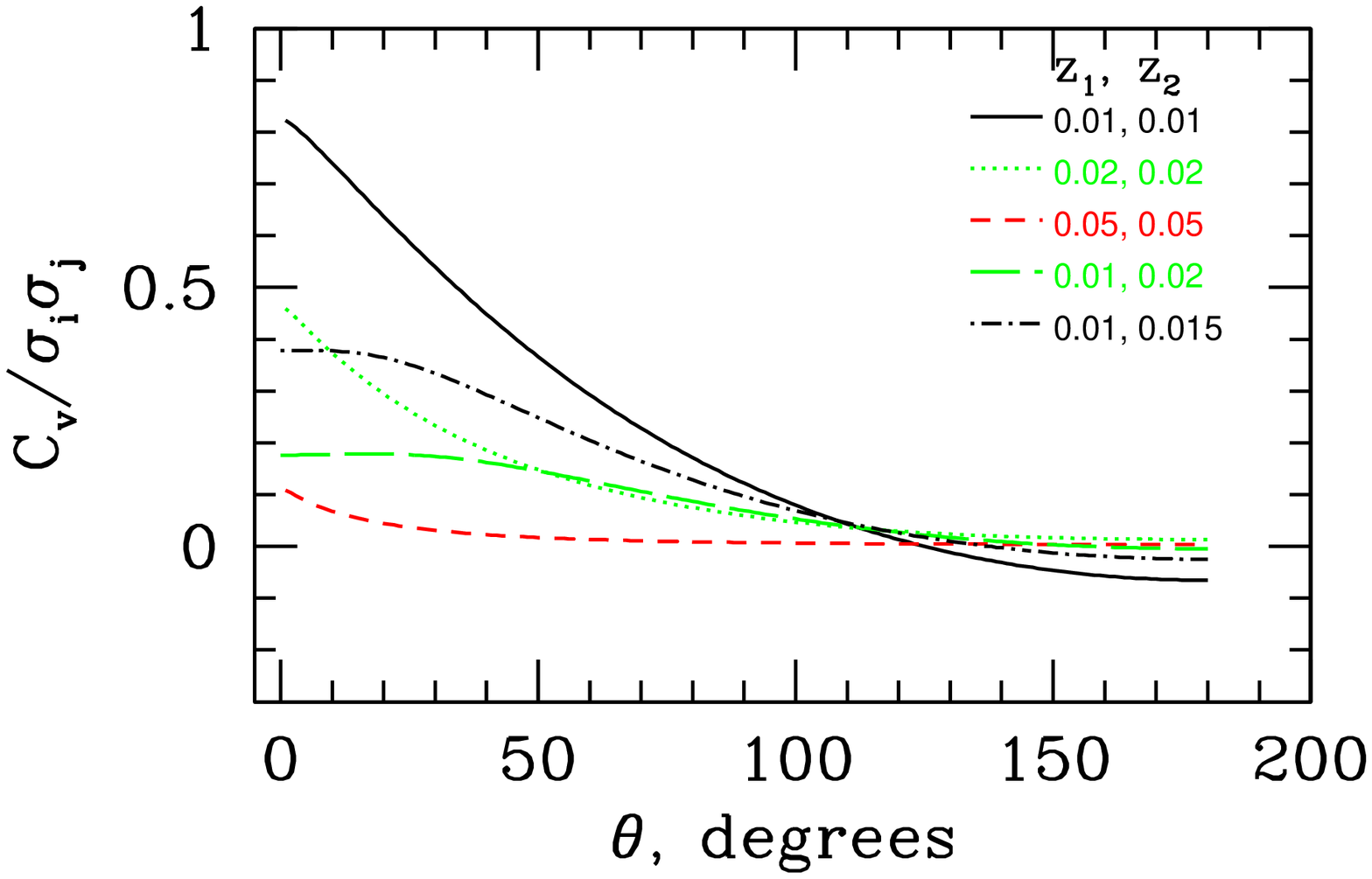}}
\centerline{\includegraphics[width=8.0cm]{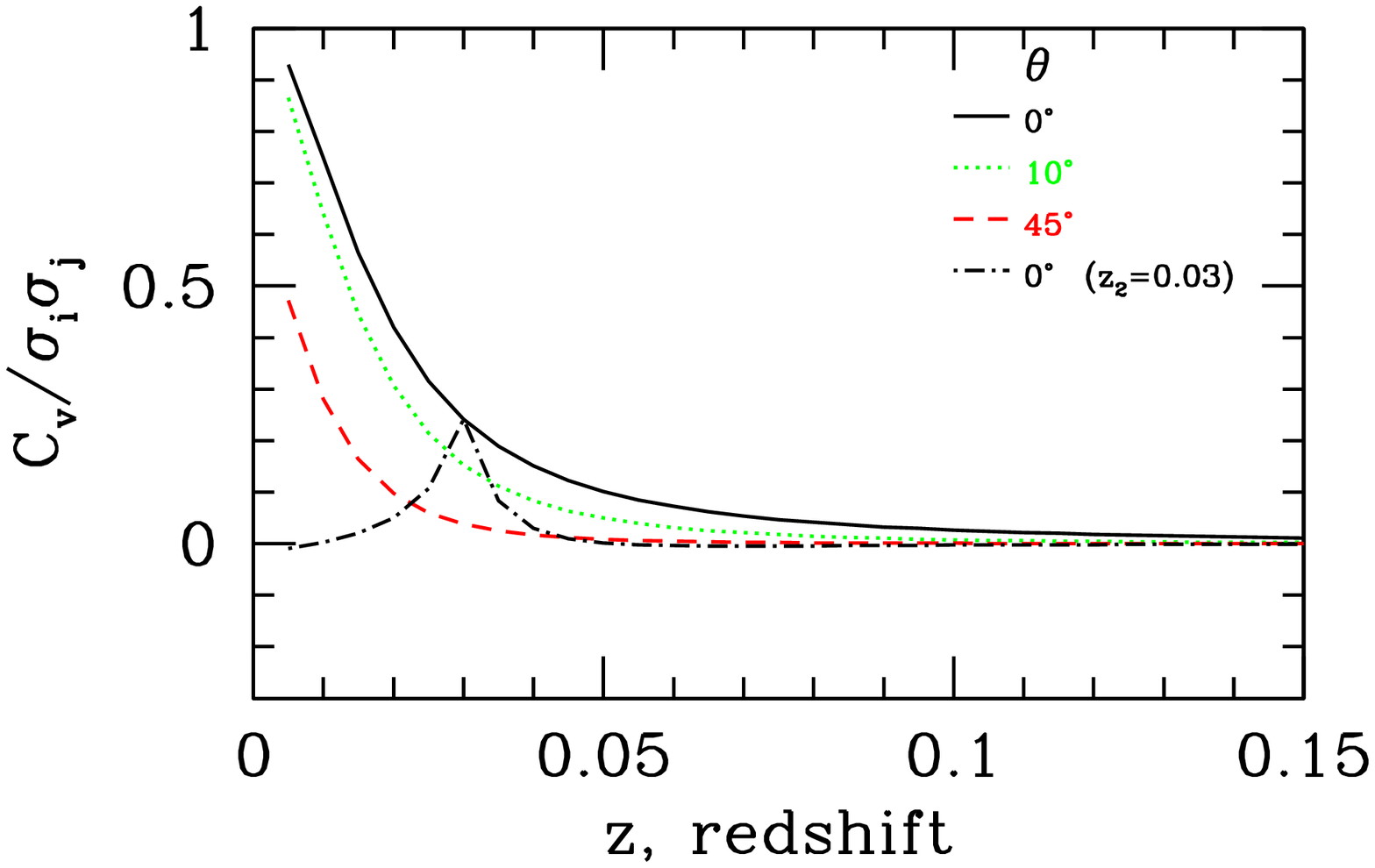}}
\caption{The ratio of the covariance from peculiar velocities $C_v$, 
compared to the random errors $\sigma$, for a pair of supernovae, over a 
range of angular separations. In the upper panel 
we vary the separation 
on the sky, $\theta$. In the lower panel we vary the redshift, 
with them both at the same $z$ or with one supernovae fixed at $z=0.03$ (dash-dot).
}
\label{errors}
\end{figure}

In Fig.~\ref{errors} we compare the covariance of $\delta d_{L}/d_L$ 
from peculiar velocities for a pair of SN
\be
C_v(i,j)=\left( 1 - \frac{(1+z)^2}{H\: d_L}\right)_i
\left( 1 - \frac{(1+z)^2}{H\: d_L}\right)_j
\xi(\bri,\brj)\label{Cv}
\ee
to the standard uncorrelated  random errors  given by 
\be
\sigma(i)^{2}=\left({\ln(10)\over 5}\right)^{2}
(\sigma_{m}^{2}+\mu_{\rm err}(i)^{2})+\left( 1 - \frac{(1+z)^2}{H\: d_L}\right)_i^2 \sigma_v^2 \label{diag}
\ee
with  $\{\Omega_m,\Omega_b,h,n_s, w,\sigma_8\}$
=\{0.3,0.05,0.7,0.96,-1,0.85\} and
$\{\mu_{err},
\sigma_m,\sigma_v\}=\{0.1,0.08,300\}$.
We see that the PV covariance is comparable with the uncorrelated  errors for 
low-$z$, and we note that the correlated errors become more significant the 
larger the dataset.

\section{Constraints from current data}\label{results}
\vspace{-0.4cm}
A uncorrelated-error-only analysis of SNe allows one to constrain the 
cosmological parameters through the Hubble parameter in (\ref{dlth}). 
Namely $\Omega_m$ and $w$, for a flat Universe. 
However, by fitting for the PV covariance 
we probe the matter power spectrum 
and can therefore constrain further parameters such as $\Omega_{b}, 
H_0, n_s, \sigma_8$, where these have their usual meaning. 
In the following analysis we also allow the SNe 
``nuisance'' parameters $M, \sigma_m, \sigma_v$ to vary. These are often set to 
fixed values, however marginalizing over them allows a better estimate of the
uncertainty in the other parameters. 
The weighted integration of the matter power spectrum in \eq{xith} has similar weights to
those used in evaluating $\sigma_{8}$. Therefore, the higher $\sigma_{8}$, the stronger the correlations between the projected velocities, obeying 
$C_v \propto \sigma_8^2$ for the other parameters fixed.

We analyse nearby supernovae ($z\leq 0.12$) from~\cite{jhariekir06} who
find improved luminosity distances to 133 supernovae from a
multicolour light curve method. Following ~\cite{jhariekir06} we
exclude 9 supernovae from the set. These are supernovae that are
unsuitable due to bad lightcurve fits, those who have their first
observation more than 20 days after maximum light, are hosted in
galaxies with excessive extinction ($A_V^0>2.0$ mag) and one outlier
(SN1999e). This leaves 124 supernovae $z\in[0.0023,0.12]$, which have an average
separation of $\bar{r}=108\: h^{-1}$Mpc, a mean redshift $\bar{z}=0.024$, 
and herein we refer to this dataset as our ``low-$z$'' SNe.


The likelihood is given by
\[
{\cal L} \propto |\Sigma |^{-1/2}{\rm e}^{-{1\over 2}{ x}^{T}\Sigma^{-1}{x}}
\]
 where $\Sigma(i,j) =C_{v}(i,j)+\sigma(i)^{2}\delta_{ij}$ and 
$x=[d_L^{obs}-d_L(z)]/d_L(z)$ with $d_L^{obs}$  given by \eq{dlobs}, and $d_L(z)$ by \eq{dlth}.
The matter power spectrum can be evaluated either numerically
(e.g.\ CAMB \cite{lewcha99}) or using 
analytical approximations \cite{eishu97}.
We assume a flat Universe with a cosmological constant $(w=-1)$, a Big
Bang Nucleosynthesis (BBN) prior $\Omega_{b} h^{2}
\sim{\mathcal N}(0.0214,0.002)$~\citep{bbn}, and a Hubble Space Telescope 
(HST) prior $h\sim {\cal N}(0.72,0.08)$~\citep{hst}. These two priors remove
models that are wildly at odds with standard cosmological probes, but
do not unduly bias results towards standard cosmology. The likelihood
has almost negligible dependence on $n_s$, and to keep it in a range
consistent with CMB and large scale structure estimates we give it a
uniform prior $n\in[-0.9,1.1]$. The other parameters are all given
broad uniform priors. We use the standard Markov Chain Monte Carlo
(MCMC) method to generate samples from the posterior distribution of
the parameters~\citep{cosmomc}.  Convergence was checked using multiple chains with
different starting positions, and also the $R-1$ statistic
~\citep{mcmc}.  We also checked that the estimated posterior
distributions reduced to the prior distributions when no data was used~\footnote{We note 
that when only weak data are present, the implied priors can
  have significant effect. In particular, the default \texttt{cosmomc}
  parametrisation is not suitable for our work in absence of CMB or
  comparably constraining datasets.}. All the analysis was checked with two completely 
independent codes and MCMC chains.

\begin{figure*}
\begin{center}
\begin{minipage}{11.5cm}
\begin{tabular}{|ll||r|r|r|r|r|}
\hline\hline
& & $w$ & $\sigma_{8}$ &
$\Omega_{m}$ & $\sigma_v$ & $\sigma_m$\\
\hline
{\bf A)} & low-$z$+PV+BBN+HST&
-1 &
$0.79^{+0.22}_{-0.22}$ &
$0.48^{+0.30}_{-0.29}$ &
$275^{+69}_{-70}$&
$0.08^{+0.03}_{-0.04}$
\\ \hline
{\bf B)} & high-$z$ &
-1 &
  &
$0.27^{+0.03}_{-0.03}$ &
$363^{+169}_{-185}$ &
$0.12^{+0.02}_{-0.02}$
\\ \hline
{\bf C)} & all-$z$+PV+BBN+HST  &
-1 &
$0.78^{+ 0.23}_{- 0.23}$&
$0.30^{+ 0.04}_{- 0.04}$&
$301^{+ 53.9}_{- 53.4}$&
$0.1^{+ 0.02}_{- 0.02}$
\\ \hline
{\bf D)} & high-$z$+WMAP &
$-0.96^{+0.09}_{-0.09}$ & 
$0.76^{+0.07}_{-0.07}$ &
$0.26^{+0.03}_{-0.03}$ & 
$191^{+97}_{-104}$
& $0.10^{+0.02}_{-0.02}$
\\ \hline
{\bf E)} & high-$z$+PV+WMAP  &
$-0.94^{+0.08}_{-0.08}$ & 
$0.75^{+0.06}_{-0.06}$ &
$0.26^{+0.02}_{-0.03}$ & 
$149^{+107}_{-108}$
& $0.11^{+0.02}_{-0.02}$
\\ \hline
{\bf F)} & all-$z$+WMAP &
$-0.93^{+0.07}_{-0.07}$ &
$0.75^{+0.06}_{-0.07}$ &
$0.26^{+0.02}_{-0.02}$ &
$395^{+42}_{-42}$ &
$0.11^{+0.02}_{-0.02}$
\\ \hline
{\bf G)} & all-$z$+PV+WMAP &
$-0.86^{+0.08}_{-0.08}$ &
$0.72^{+0.06}_{-0.07}$ &
$0.28^{+0.03}_{-0.03}$ &
$292^{+44}_{-45}$ &
$0.11^{+0.02}_{-0.02}$
\\ \hline
{\bf H)} & WMAP only                                &
$-0.99^{+0.22}_{-0.22}$ &
$0.76^{+0.09}_{-0.09}$ &
$0.25^{+0.05}_{-0.05}$& &
 \\ \hline\hline
\end{tabular}
\tabcaption{
Mean and 68\% confidence limits on the cosmological parameters, and $\sigma_v, \sigma_m$, 
using different
combinations of the WMAP and SNe datasets, with and without including the 
peculiar velocity covariance matrix (PV). See text for discussion. }
\label{tab}
\end{minipage}
\hfill
\begin{minipage}{6.0cm}
\includegraphics[height=4.0cm]{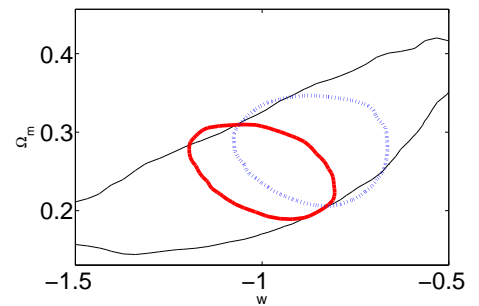}
\caption{95\% confidence limits on $\Omega_{m}$ and $w$, for WMAP only (thin black), 
WMAP with high-$z$ SNe (thick red), and WMAP with all-$z$ SNe including PVs (dotted blue)
}\label{w}
\end{minipage}
\end{center}
\end{figure*}

The low-$z$ results are given in row A of Table~\ref{tab}. 
As a non-zero $\sigma_{8}$ is needed for the velocity correlations,
these results indicate the correlations are detected at the 3.6$\sigma$
level.
We also perform the MCMC runs without including the PV covariance 
matrix $C_v$, and 
we find $-2 \ln{\cal L}_{\rm max}$ (where ${\cal L}$ is the
likelihood) increases by 19.3. As the likelihood no longer depends
on $\{\sigma_8,\Omega_b,h,n_s\}$ we have effectively removed four 
parameters. 

When estimating the cosmological parameters from SNe, a low redshift cut is 
usually imposed, to reduce the
effects of the PVs. For example in~\cite{davis07}, their
SNe dataset have 192 SNe with $z\in[0.016,1.76]$ and a mean redshift of
$\bar{z}=0.48$. Herein we refer to this dataset as our ``high-$z$'' SNe. 
Our MCMC results for just this data (without including 
PVs) are shown in row B of Table~\ref{tab}. Although we marginalize over the SNe parameters
$\{M,\sigma_{v},\sigma_{s}\}$, 
our constraints on $\Omega_{m}$ are still in excellent agreement with those 
obtained in~\cite{davis07}.

We now combine the low-$z$ and the high-$z$ datasets, to make an ``all-$z$'' dataset. 
We used the overlapping SNe in the two datasets to
estimate a small normalizing offset to the magnitudes from the latter data set, 
(the extra magnitude error is negligibly small). The same procedure was used by~\cite{davis07}
in constructing their dataset. After eliminating duplicated SNe, our
combined all-$z$ dataset has 271 SNe with $z\in[0.0023,1.76]$, and
$\bar{z}=0.35$. Our all-$z$ results are given in row
C of Table~\ref{tab}. The constraints on
$\sigma_{8}$ broaden slightly due to a mild degeneracy between
$\sigma_{8}$ and $\Omega_{m}$ which is broken by the addition of the
higher $z$ SNe, but pushes $\sigma_8$ to the region of higher
uncertainty. 
The increase in the minimum of $-2\ln {\cal L}$, when
$\sigma_{8}$ is set to zero, is now 16.4. 

One the main aims of cosmology is to test the cosmological constant
prediction $w=-1$. Due to a degeneracy between $w$ and $\Omega_{m}$,
it is necessary to combine the SNe with another data source. Here we
use the WMAP CMB data~\cite{WMAP06}. We continue to assume a flat
Universe, and we now allow $w$ to be a free parameter.
The results are given in rows D to H of Table~\ref{tab} and in
Fig.~\ref{w}.
As can be seen by comparing rows D and E, if a redshift cutoff of $z\ge 0.016$ is made, then current data has a systematic error of $\delta w=0.02$ when PVs are not included. This is several times smaller than the statistical error of $\delta w=0.08$. 
 However, comparing rows F and  G, shows that if no redshift cutoff is made,
neglecting the correlated PVs results in a systematic error of
$\delta w=0.07$ which is about as large as the statistical error.


\section{Forecasts}\label{FC}
\vspace{-0.4cm}
We now consider the relevance of peculiar velocities to future supernovae 
surveys, using a Fisher matrix analysis. 
The Fisher (information) matrix probes the ability of an experiment to constrain 
parameters, by looking at the dependence of the likelihood:
${\rF}_{\alpha \beta}
\equiv - \left<\frac{\partial^2 \ln{\mathcal L}}{\partial p_\alpha \partial 
p_\beta}\right> $
\bea
{\rF}_{\alpha \beta}&=& \bd,_{\alpha}{\rm C}^{-1}\bd^{\rm T},_\beta
 + \frac{1}{2}\:{\rm Tr}\left({\rm C^{-1}C,_\alpha C^{-1}C,_\beta}\right)
\eea
Often the second term is ignored, however 
we find that this approximation is no longer valid when including PVs.
Unmarginalised $1\sigma$ errors on parameter $p_\alpha$ are given by 
$\sqrt{(1/\rF_{\alpha\alpha})}$, 
and the equivalent marginalised errors by $\sqrt{(\{\rF^{-1}\}_{\alpha\alpha})}$. 
Generally, the 
inverted Fisher matrix $\rF^{-1}$ provides the expected covariance matrix for the 
parameters.


We consider two future supernoave experiements that aim to detect high and 
low-redshift supernoave respectively: the Supernova/Acceleration Probe 
(SNAP,~\citep{SNAP2}), and the Nearby Supernova Factory (SNfactory,~\citep{SNfact}).
The baseline SNfactory program is to obtain spectrophotometric lightcurves for 
SNe in the redshift range $0.03<z<0.08$, with the assumption that these SNe are far 
enough away for correlated peculiar velocities to not contribute significantly to 
the error budget.
However, in Fig~\ref{errors} we can see the contribution of correlated PVs is not irrelevant for 
$z \sim 0.03$, and as the number of SNe increases the overall uncertainty 
from random errors 
(such as intrinsic magnitude scattter and instrumental noise) gets beaten down by 
a $1/\sqrt{N}$ factor, while the correlated errors from coherent peculiar velocities 
do not.
Thus at any redshift, these peculiar velocity errors will begin to dominate for 
some large number of SNe. We investigate the situation for the SNfactory by 
considering 300 SNe randomly distributed over a rectangular area of 10,000 sq 
degrees, and redshifts $0.03 < z < 0.08$.
We also include high-$z$ from a SNAP-like survey, and model this as 2000 SNe 
randomly distributed over 10 sq degrees with $0.2 < z < 1.7$, and we assume 
PVs are irrelevant for these SNe.
We take our fiducial model to be a flat $\Lambda$CDM cosmology with 
$\Omega_m$=0.3, $\Omega_b=0.05$, $h=0.7$, $n_s=0.96$, $w=-1$, $\sigma_8=0.85$, 
and nuisance parameters $\sigma_v=300$ km/s, $\sigma_m$=0.1, $\mu_{err}=0.1$. 
We further marginalise over $M$.

\begin{figure}
\begin{centering}
\includegraphics[width=7.0cm]{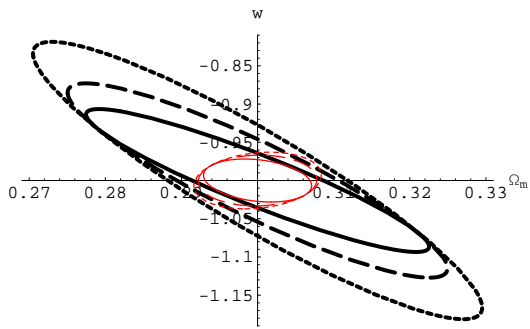}
\end{centering}
\figcaption{The marginalised $1\sigma$ contours for $\Omega_m$ and $w$ from a SNAP-like 
high-$z$ SNe survey in a flat $\Lambda$CDM cosmology (short dashed lines). We also consider 
including 300 low-$z$ SNe from a SNfactory-like survey, while ignoring peculiar 
velocities (solid lines) and including them (long dashed lines). Smaller red contours 
include cosmic variance limited CMB, up to $\ell=2000$.}
\label{SNcon}
\end{figure}

In Fig~\ref{SNcon} we compare the marginalised $1\sigma$ contours obtained  when 
the coherent PVs are ingnored and included, for our hypothetical 
SNAP and SNfactory surveys.
We see that, even for a cut of $z > 0.03$ the error bars on $\Omega_m$ and $w$ 
will be considerably underestimated if the peculiar velocites are ignored, and in 
particular the marginalised error on $w$ increases by 35\% from 0.062 to 0.084, for the 
SNe alone.
We therefore conclude that it is essential to include a full covariance matrix 
analysis at these redshifts, to avoid significantly underestimating the errors.
That the error bars increase rather than decrease indicates that the extra 
information avaliable in the peculiar velocities is outweighed by the extra parameter 
space $\{\Omega_b, h, n_s, \sigma_8\}$.

In light of the extra effort required for the low-$z$ SNe we consider excluding them. 
However in Fig~\ref{SNcon} we see that the contours increase significantly 
when low-$z$ SNe are not available (marginalised error on $w$ increase to 0.12), 
even when CMB data is included.
Therefore we conclude that low-$z$ SNe play a vital role in constraining the 
cosmological parameters, but a full covariance matrix analysis must be done.

\begin{figure}
\begin{centering}
\includegraphics[width=7.0cm]{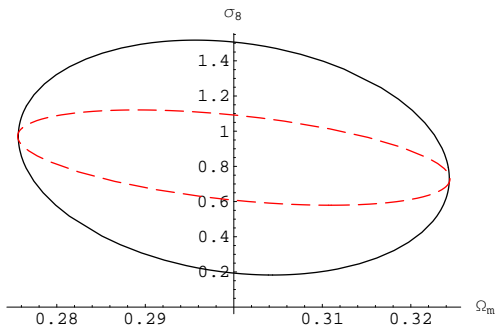}
\end{centering}
\figcaption{The marginalised $1\sigma$ contours for $\Omega_m$ and $\sigma_8$, from 
1000 low-$z$ SNe in a $\Lambda$CDM cosmology, with $0.03<z<0.08$ (black solid line) 
and $z<0.08$ (red dashed line), combined with a SNAP-like high-$z$ SNe survey.}
\label{SN1000}
\end{figure}

We have shown that a low redshift cut of 0.03 is futile in avoiding peculiar 
velocites when we have 100's of SNe.
We therefore consider that no lower bound on redshift is needed at all, and we 
extend our study to include SNe with $z<0.03$.
We consider a survey of 1000 SNe randomly distributed over 10,000 sq degrees and 
redshifts $z<0.08$, 
and compare it to a similar survey of 1000 SNe, but with $0.03 < z <0.08$. 
We combine them with a high-$z$ SNAP-like survey as above.
In Fig~\ref{SN1000} we see that by including very low-$z$ SNe ($z<0.03$) we have 
gained free information about $\sigma_8$ - `free' because the error bars on 
$\Omega_m$ (and $w$) stay the same. However, a
lower bound on $z$ may be useful in reducing the effects of non-linearities.

\section{Discussion}
We have found that when no redshift cutoff is imposed, the current SNe data detects correlations in PVs at   the  $3.6\sigma$ level. We also  found that the current constraints on $w$ are not sensitive to the PVs if SNe with $z<0.016$ are excluded. However, we have shown that future large data sets will be sensitive to the PVs even if a lower bound of  $z\ge0.03$ is used.   

An alternative approach of accounting for the correlated peculiar velocities is to estimate the underlying density field from galaxy redshift surveys and then use this to try and remove  peculiar velocity at each SNe \citep{nehuco07}. 
As this 
method has different systematics to the statistical modeling method we have investigated, we believe it will be a useful cross-check  to apply both methods separately to future SNe data sets and compare the results.

\vspace{0.5cm}
\centerline{\bf Appendix}

\renewcommand{\theequation}{A\arabic{equation}}
\setcounter{equation}{0}


For perturbations, $\rho(1+\delta)$, and peculiar velocity 
$\bv$ we find the perturbed Fourier space continuity equation, in co-moving coordinates, is 
$
\delta_{\bk}' -i\bk\cdot\bv_\bk=0
$
where prime indicates differentiation with respect to conformal time, $\eta$.
Using the linear approximation $\delta_\bk(\eta)=D(\eta)\tilde{\delta}_\bk$, and assuming 
that the Universe has no vorticity $(\nabla \times \bv=0)$ leads to 
$
\bv_\bk 
=-iD'\frac{\tilde{\delta}_\bk}{k^2}\bk,
$
where $D(z)$ is the growth function. Therefore, the Fourier space correlation between the $i$th component of the velocity field at time $\eta_{A}$ and the $j$th component of the velocity field at time $\eta_{B}$ is given by
\[
\left<\rv^i_{\bf k_{A}} \rv^{j*}_{\bf k_{B}} \right> = D'(\eta_{A})D'(\eta_{B})(2\pi)^3
\delta({\bf k_{A}}-{\bf k_{B}})P(k_{A})\frac{k_{A}^i k_{A}^j}{(k_{A})^4},
\]
where $\left<\delta_{\bf k_{A}} \delta_{\bf k_{B}} \right>=(2\pi)^3\delta({\bf k_{A}}-{\bf k_{B}})P(k_{A})$, and the subscripts denote the quantity at time $\eta_{A}$ or $\eta_{B}$. 
This corrects equation (B1) of~\cite{Monteagudo:2005ys}, and our resulting \eq{Cv} 
for $C_v$  corrects those of~\cite{Cooray:2006ft}, among others.

\centerline{---------------------------------------------------}


We thank Joe Silk, Pedro Ferreira, and Uro\v{s} Seljak for useful 
conversations. CG is funded by the Beecroft Institute for Particle Astrophysics and 
Cosmology, KRL by Glasstone research 
fellowship and Christ Church college, AS by Oxford Astrophysics.
Results were computed on the UK-CCC COSMOS supercomputer.


\bibliography{bib}

\begin{thebibliography}{41}
\expandafter\ifx\csname natexlab\endcsname\relax\def\natexlab#1{#1}\fi
\expandafter\ifx\csname bibnamefont\endcsname\relax
  \def\bibnamefont#1{#1}\fi
\expandafter\ifx\csname bibfnamefont\endcsname\relax
  \def\bibfnamefont#1{#1}\fi
\expandafter\ifx\csname citenamefont\endcsname\relax
  \def\citenamefont#1{#1}\fi
\expandafter\ifx\csname url\endcsname\relax
  \def\url#1{\texttt{#1}}\fi
\expandafter\ifx\csname urlprefix\endcsname\relax\def\urlprefix{URL }\fi
\providecommand{\bibinfo}[2]{#2}
\providecommand{\eprint}[2][]{\url{#2}}

\bibitem[{\citenamefont{Riess et~al.}(1998)}]{Riess:1998cb}
\bibinfo{author}{\bibfnamefont{A.~G.} \bibnamefont{Riess}} \bibnamefont{et~al.}
  (\bibinfo{collaboration}{Supernova Search Team}), \bibinfo{journal}{Astron.
  J.} \textbf{\bibinfo{volume}{116}}, \bibinfo{pages}{1009}
  (\bibinfo{year}{1998}), \eprint{astro-ph/9805201}.

\bibitem[{\citenamefont{Perlmutter et~al.}(1999)}]{Perlmutter:1998np}
\bibinfo{author}{\bibfnamefont{S.}~\bibnamefont{Perlmutter}}
  \bibnamefont{et~al.} (\bibinfo{collaboration}{Supernova Cosmology Project}),
  \bibinfo{journal}{Astrophys. J.} \textbf{\bibinfo{volume}{517}},
  \bibinfo{pages}{565} (\bibinfo{year}{1999}), \eprint{astro-ph/9812133}.

\bibitem[{\citenamefont{Astier et~al.}(2006)}]{Astier:2005qq}
\bibinfo{author}{\bibfnamefont{P.}~\bibnamefont{Astier}} \bibnamefont{et~al.},
  \bibinfo{journal}{Astron. Astrophys.} \textbf{\bibinfo{volume}{447}},
  \bibinfo{pages}{31} (\bibinfo{year}{2006}).

\bibitem[{\citenamefont{Riess et~al.}(2006)}]{Riess:2006fw}
\bibinfo{author}{\bibfnamefont{A.~G.} \bibnamefont{Riess}} \bibnamefont{et~al.}
  (\bibinfo{year}{2006}), \eprint{astro-ph/0611572}.

\bibitem[{\citenamefont{Wood-Vasey et~al.}(2007)}]{essence}
\bibinfo{author}{\bibfnamefont{W.~M.} \bibnamefont{Wood-Vasey}}
  \bibnamefont{et~al.} (\bibinfo{year}{2007}), \eprint{astro-ph/0701041}.

\bibitem[{\citenamefont{{Aldering} et~al.}(2002)}]{SNfact}
\bibinfo{author}{\bibfnamefont{G.}~\bibnamefont{{Aldering}}}
  \bibnamefont{et~al.}, in \emph{\bibinfo{booktitle}{Survey and Other Telescope
  Technologies and Discoveries}}, edited by
  \bibinfo{editor}{\bibfnamefont{J.~A.} \bibnamefont{{Tyson}}}
  \bibnamefont{and} \bibinfo{editor}{\bibfnamefont{S.}~\bibnamefont{{Wolff}}}
  (\bibinfo{year}{2002}), vol. \bibinfo{volume}{4836}, pp.
  \bibinfo{pages}{61--72}.

\bibitem[{\citenamefont{{Dilday} et~al.}(2005)}]{sdsssn}
\bibinfo{author}{\bibfnamefont{B.}~\bibnamefont{{Dilday}}}
  \bibnamefont{et~al.}, in \emph{\bibinfo{booktitle}{Bulletin of the American
  Astronomical Society}} (\bibinfo{year}{2005}), vol.~\bibinfo{volume}{37}, pp.
  \bibinfo{pages}{1459--+}.

\bibitem[{\citenamefont{Hamuy et~al.}(2005)}]{carnegysn}
\bibinfo{author}{\bibfnamefont{M.}~\bibnamefont{Hamuy}} \bibnamefont{et~al.}
  (\bibinfo{year}{2005}), \eprint{astro-ph/0512039}.

\bibitem[{\citenamefont{Aldering}(2005)}]{SNAP2}
\bibinfo{author}{\bibfnamefont{G.}~\bibnamefont{Aldering}},
  \bibinfo{journal}{New Astron. Rev.} \textbf{\bibinfo{volume}{49}},
  \bibinfo{pages}{346} (\bibinfo{year}{2005}).

\bibitem[{\citenamefont{Riess et~al.}(1995)\citenamefont{Riess, Press, and
  Kirshner}}]{rieprekir95}
\bibinfo{author}{\bibfnamefont{A.~G.} \bibnamefont{Riess}},
  \bibinfo{author}{\bibfnamefont{W.~H.} \bibnamefont{Press}}, \bibnamefont{and}
  \bibinfo{author}{\bibfnamefont{R.~P.} \bibnamefont{Kirshner}},
  \bibinfo{journal}{Astrophys. J.} \textbf{\bibinfo{volume}{445}},
  \bibinfo{pages}{L91} (\bibinfo{year}{1995}), \eprint{astro-ph/9412017}.

\bibitem[{\citenamefont{Riess et~al.}(1997)\citenamefont{Riess, Davis, Baker,
  and Kirshner}}]{riess97}
\bibinfo{author}{\bibfnamefont{A.~G.} \bibnamefont{Riess}},
  \bibinfo{author}{\bibfnamefont{M.}~\bibnamefont{Davis}},
  \bibinfo{author}{\bibfnamefont{J.}~\bibnamefont{Baker}}, \bibnamefont{and}
  \bibinfo{author}{\bibfnamefont{R.~P.} \bibnamefont{Kirshner}}
  (\bibinfo{year}{1997}), \eprint{astro-ph/9707261}.

\bibitem[{\citenamefont{Zehavi et~al.}(1998)\citenamefont{Zehavi, Riess,
  Kirshner, and Dekel}}]{zehriekir98}
\bibinfo{author}{\bibfnamefont{I.}~\bibnamefont{Zehavi}},
  \bibinfo{author}{\bibfnamefont{A.~G.} \bibnamefont{Riess}},
  \bibinfo{author}{\bibfnamefont{R.~P.} \bibnamefont{Kirshner}},
  \bibnamefont{and} \bibinfo{author}{\bibfnamefont{A.}~\bibnamefont{Dekel}},
  \bibinfo{journal}{Astrophys. J.} \textbf{\bibinfo{volume}{503}},
  \bibinfo{pages}{483} (\bibinfo{year}{1998}), \eprint{astro-ph/9802252}.

\bibitem[{\citenamefont{{Bonacic} et~al.}(2000)\citenamefont{{Bonacic},
  {Schommer}, {Suntzeff}, and {Phillips}}}]{bonacic00}
\bibinfo{author}{\bibfnamefont{A.}~\bibnamefont{{Bonacic}}},
  \bibinfo{author}{\bibfnamefont{R.~A.} \bibnamefont{{Schommer}}},
  \bibinfo{author}{\bibfnamefont{N.~B.} \bibnamefont{{Suntzeff}}},
  \bibnamefont{and} \bibinfo{author}{\bibfnamefont{M.~M.}
  \bibnamefont{{Phillips}}}, \bibinfo{journal}{Bulletin of the American
  Astronomical Society} \textbf{\bibinfo{volume}{32}}, \bibinfo{pages}{1285}
  (\bibinfo{year}{2000}).

\bibitem[{\citenamefont{Radburn-Smith et~al.}(2004)\citenamefont{Radburn-Smith,
  Lucey, and Hudson}}]{radluchud04}
\bibinfo{author}{\bibfnamefont{D.~J.} \bibnamefont{Radburn-Smith}},
  \bibinfo{author}{\bibfnamefont{J.~R.} \bibnamefont{Lucey}}, \bibnamefont{and}
  \bibinfo{author}{\bibfnamefont{M.~J.} \bibnamefont{Hudson}}
  (\bibinfo{year}{2004}), \eprint{astro-ph/0409551}.

\bibitem[{\citenamefont{Bonvin et~al.}(2006{\natexlab{a}})\citenamefont{Bonvin,
  Durrer, and Kunz}}]{bonvin}
\bibinfo{author}{\bibfnamefont{C.}~\bibnamefont{Bonvin}},
  \bibinfo{author}{\bibfnamefont{R.}~\bibnamefont{Durrer}}, \bibnamefont{and}
  \bibinfo{author}{\bibfnamefont{M.}~\bibnamefont{Kunz}},
  \bibinfo{journal}{Phys. Rev. Lett.} \textbf{\bibinfo{volume}{96}},
  \bibinfo{pages}{191302} (\bibinfo{year}{2006}{\natexlab{a}}),
  \eprint{astro-ph/0603240}.

\bibitem[{\citenamefont{Haugboelle et~al.}(2006)}]{haugboelle06}
\bibinfo{author}{\bibfnamefont{T.}~\bibnamefont{Haugboelle}}
  \bibnamefont{et~al.} (\bibinfo{year}{2006}), \eprint{astro-ph/0612137}.

\bibitem[{\citenamefont{Jha et~al.}(2007)\citenamefont{Jha, Riess, and
  Kirshner}}]{jhariekir06}
\bibinfo{author}{\bibfnamefont{S.}~\bibnamefont{Jha}},
  \bibinfo{author}{\bibfnamefont{A.~G.} \bibnamefont{Riess}}, \bibnamefont{and}
  \bibinfo{author}{\bibfnamefont{R.~P.} \bibnamefont{Kirshner}},
  \bibinfo{journal}{Astrophys. J.} \textbf{\bibinfo{volume}{659}},
  \bibinfo{pages}{122} (\bibinfo{year}{2007}), \eprint{astro-ph/0612666}.

\bibitem[{\citenamefont{Watkins and Feldman}(2007)}]{watfeld07}
\bibinfo{author}{\bibfnamefont{R.}~\bibnamefont{Watkins}} \bibnamefont{and}
  \bibinfo{author}{\bibfnamefont{H.~A.} \bibnamefont{Feldman}}
  (\bibinfo{year}{2007}), \eprint{astro-ph/0702751}.

\bibitem[{\citenamefont{Conley et~al.}(2007)}]{conley07}
\bibinfo{author}{\bibfnamefont{A.}~\bibnamefont{Conley}} \bibnamefont{et~al.}
  (\bibinfo{year}{2007}), \eprint{arXiv:0705.0367 [astro-ph]}.

\bibitem[{\citenamefont{Wang}(2007)}]{wang07}
\bibinfo{author}{\bibfnamefont{L.}~\bibnamefont{Wang}} (\bibinfo{year}{2007}),
  \eprint{arXiv:0705.0368 [astro-ph]}.

\bibitem[{\citenamefont{Neill et~al.}(2007)\citenamefont{Neill, Hudson, and
  Conley}}]{nehuco07}
\bibinfo{author}{\bibfnamefont{J.~D.} \bibnamefont{Neill}},
  \bibinfo{author}{\bibfnamefont{M.~J.} \bibnamefont{Hudson}},
  \bibnamefont{and} \bibinfo{author}{\bibfnamefont{A.}~\bibnamefont{Conley}}
  (\bibinfo{year}{2007}), \eprint{arXiv:0704.1654 [astro-ph]}.

\bibitem[{\citenamefont{Hannestad et~al.}(2007)\citenamefont{Hannestad,
  Haugboelle, and Thomsen}}]{Han}
\bibinfo{author}{\bibfnamefont{S.}~\bibnamefont{Hannestad}},
  \bibinfo{author}{\bibfnamefont{T.}~\bibnamefont{Haugboelle}},
  \bibnamefont{and} \bibinfo{author}{\bibfnamefont{B.}~\bibnamefont{Thomsen}}
  (\bibinfo{year}{2007}), \eprint{arXiv:0705.0979 [astro-ph]}.

\bibitem[{\citenamefont{Hui and Greene}(2006)}]{Hui:2005nm}
\bibinfo{author}{\bibfnamefont{L.}~\bibnamefont{Hui}} \bibnamefont{and}
  \bibinfo{author}{\bibfnamefont{P.~B.} \bibnamefont{Greene}},
  \bibinfo{journal}{Phys. Rev.} \textbf{\bibinfo{volume}{D73}},
  \bibinfo{pages}{123526} (\bibinfo{year}{2006}), \eprint{astro-ph/0512159}.

\bibitem[{\citenamefont{Sasaki}(1987)}]{sasaki87}
\bibinfo{author}{\bibfnamefont{M.}~\bibnamefont{Sasaki}},
  \bibinfo{journal}{Mon. Not. Roy. Astron. Soc.}
  \textbf{\bibinfo{volume}{228}}, \bibinfo{pages}{653} (\bibinfo{year}{1987}).

\bibitem[{\citenamefont{{Sugiura} et~al.}(1999)\citenamefont{{Sugiura},
  {Sugiyama}, and {Sasaki}}}]{sugsugsas99}
\bibinfo{author}{\bibfnamefont{N.}~\bibnamefont{{Sugiura}}},
  \bibinfo{author}{\bibfnamefont{N.}~\bibnamefont{{Sugiyama}}},
  \bibnamefont{and} \bibinfo{author}{\bibfnamefont{M.}~\bibnamefont{{Sasaki}}},
  \bibinfo{journal}{Progress of Theoretical Physics}
  \textbf{\bibinfo{volume}{101}}, \bibinfo{pages}{903} (\bibinfo{year}{1999}).

\bibitem[{\citenamefont{Pyne and Birkinshaw}(2004)}]{pynbirk95}
\bibinfo{author}{\bibfnamefont{T.}~\bibnamefont{Pyne}} \bibnamefont{and}
  \bibinfo{author}{\bibfnamefont{M.}~\bibnamefont{Birkinshaw}},
  \bibinfo{journal}{Mon. Not. Roy. Astron. Soc.}
  \textbf{\bibinfo{volume}{348}}, \bibinfo{pages}{581} (\bibinfo{year}{2004}),
  \eprint{astro-ph/0310841}.

\bibitem[{\citenamefont{Bonvin et~al.}(2006{\natexlab{b}})\citenamefont{Bonvin,
  Durrer, and Gasparini}}]{bondurgas05}
\bibinfo{author}{\bibfnamefont{C.}~\bibnamefont{Bonvin}},
  \bibinfo{author}{\bibfnamefont{R.}~\bibnamefont{Durrer}}, \bibnamefont{and}
  \bibinfo{author}{\bibfnamefont{M.~A.} \bibnamefont{Gasparini}},
  \bibinfo{journal}{Phys. Rev.} \textbf{\bibinfo{volume}{D73}},
  \bibinfo{pages}{023523} (\bibinfo{year}{2006}{\natexlab{b}}),
  \eprint{astro-ph/0511183}.

\bibitem[{\citenamefont{Gorski}(1988)}]{Gorski}
\bibinfo{author}{\bibfnamefont{K.}~\bibnamefont{Gorski}},
  \bibinfo{journal}{Astrophys. J.} \textbf{\bibinfo{volume}{332}},
  \bibinfo{pages}{L7} (\bibinfo{year}{1988}).

\bibitem[{\citenamefont{Groth et~al.}(1989)\citenamefont{Groth, Juszkiewicz,
  and Ostriker}}]{Groth}
\bibinfo{author}{\bibfnamefont{E.~J.} \bibnamefont{Groth}},
  \bibinfo{author}{\bibfnamefont{R.}~\bibnamefont{Juszkiewicz}},
  \bibnamefont{and} \bibinfo{author}{\bibfnamefont{J.~P.}
  \bibnamefont{Ostriker}}, \bibinfo{journal}{Astrophys. J.}
  \textbf{\bibinfo{volume}{346}}, \bibinfo{pages}{558} (\bibinfo{year}{1989}).

\bibitem[{\citenamefont{{Dodelson}}(2003)}]{Scott}
\bibinfo{author}{\bibfnamefont{S.}~\bibnamefont{{Dodelson}}},
  \emph{\bibinfo{title}{{Modern cosmology}}} (\bibinfo{publisher}{Academic
  Press}, \bibinfo{year}{2003}).

\bibitem[{\citenamefont{Hernandez-Monteagudo
  et~al.}(2006)\citenamefont{Hernandez-Monteagudo, Verde, Jimenez, and
  Spergel}}]{Monteagudo:2005ys}
\bibinfo{author}{\bibfnamefont{C.}~\bibnamefont{Hernandez-Monteagudo}},
  \bibinfo{author}{\bibfnamefont{L.}~\bibnamefont{Verde}},
  \bibinfo{author}{\bibfnamefont{R.}~\bibnamefont{Jimenez}}, \bibnamefont{and}
  \bibinfo{author}{\bibfnamefont{D.~N.} \bibnamefont{Spergel}},
  \bibinfo{journal}{Astrophys. J.} \textbf{\bibinfo{volume}{643}},
  \bibinfo{pages}{598} (\bibinfo{year}{2006}), \eprint{astro-ph/0511061}.

\bibitem[{\citenamefont{Cooray and Caldwell}(2006)}]{Cooray:2006ft}
\bibinfo{author}{\bibfnamefont{A.}~\bibnamefont{Cooray}} \bibnamefont{and}
  \bibinfo{author}{\bibfnamefont{R.~R.} \bibnamefont{Caldwell}},
  \bibinfo{journal}{Phys. Rev.} \textbf{\bibinfo{volume}{D73}},
  \bibinfo{pages}{103002} (\bibinfo{year}{2006}), \eprint{astro-ph/0601377}.

\bibitem[{\citenamefont{Silberman et~al.}(2001)\citenamefont{Silberman, Dekel,
  Eldar, and Zehavi}}]{silberman01}
\bibinfo{author}{\bibfnamefont{L.}~\bibnamefont{Silberman}},
  \bibinfo{author}{\bibfnamefont{A.}~\bibnamefont{Dekel}},
  \bibinfo{author}{\bibfnamefont{A.}~\bibnamefont{Eldar}}, \bibnamefont{and}
  \bibinfo{author}{\bibfnamefont{I.}~\bibnamefont{Zehavi}},
  \bibinfo{journal}{Astrophys. J.} \textbf{\bibinfo{volume}{557}},
  \bibinfo{pages}{102} (\bibinfo{year}{2001}), \eprint{astro-ph/0101361}.

\bibitem[{\citenamefont{Lewis et~al.}(2000)\citenamefont{Lewis, Challinor, and
  Lasenby}}]{lewcha99}
\bibinfo{author}{\bibfnamefont{A.}~\bibnamefont{Lewis}},
  \bibinfo{author}{\bibfnamefont{A.}~\bibnamefont{Challinor}},
  \bibnamefont{and} \bibinfo{author}{\bibfnamefont{A.}~\bibnamefont{Lasenby}},
  \bibinfo{journal}{Astrophys. J.} \textbf{\bibinfo{volume}{538}},
  \bibinfo{pages}{473} (\bibinfo{year}{2000}), \eprint{astro-ph/9911177}.

\bibitem[{\citenamefont{Eisenstein and Hu}(1998)}]{eishu97}
\bibinfo{author}{\bibfnamefont{D.~J.} \bibnamefont{Eisenstein}}
  \bibnamefont{and} \bibinfo{author}{\bibfnamefont{W.}~\bibnamefont{Hu}},
  \bibinfo{journal}{Astrophys. J.} \textbf{\bibinfo{volume}{496}},
  \bibinfo{pages}{605} (\bibinfo{year}{1998}), \eprint{astro-ph/9709112}.

\bibitem[{\citenamefont{Kirkman et~al.}(2003)\citenamefont{Kirkman, Tytler,
  Suzuki, O'Meara, and Lubin}}]{bbn}
\bibinfo{author}{\bibfnamefont{D.}~\bibnamefont{Kirkman}},
  \bibinfo{author}{\bibfnamefont{D.}~\bibnamefont{Tytler}},
  \bibinfo{author}{\bibfnamefont{N.}~\bibnamefont{Suzuki}},
  \bibinfo{author}{\bibfnamefont{J.~M.} \bibnamefont{O'Meara}},
  \bibnamefont{and} \bibinfo{author}{\bibfnamefont{D.}~\bibnamefont{Lubin}},
  \bibinfo{journal}{Astrophys. J. Suppl.} \textbf{\bibinfo{volume}{149}},
  \bibinfo{pages}{1} (\bibinfo{year}{2003}), \eprint{astro-ph/0302006}.

\bibitem[{\citenamefont{Freedman et~al.}(2001)}]{hst}
\bibinfo{author}{\bibfnamefont{W.~L.} \bibnamefont{Freedman}}
  \bibnamefont{et~al.}, \bibinfo{journal}{Astrophys. J.}
  \textbf{\bibinfo{volume}{553}}, \bibinfo{pages}{47} (\bibinfo{year}{2001}),
  \eprint{astro-ph/0012376}.

\bibitem[{\citenamefont{{Lewis} and {Bridle}}(2002)}]{cosmomc}
\bibinfo{author}{\bibfnamefont{A.}~\bibnamefont{{Lewis}}} \bibnamefont{and}
  \bibinfo{author}{\bibfnamefont{S.}~\bibnamefont{{Bridle}}},
  \bibinfo{journal}{\prd} \textbf{\bibinfo{volume}{66}},
  \bibinfo{pages}{103511} (\bibinfo{year}{2002}).

\bibitem[{\citenamefont{Gelman and Rubin}(1992)}]{mcmc}
\bibinfo{author}{\bibfnamefont{A.}~\bibnamefont{Gelman}} \bibnamefont{and}
  \bibinfo{author}{\bibfnamefont{D.}~\bibnamefont{Rubin}},
  \bibinfo{journal}{Stat. Sci.} \textbf{\bibinfo{volume}{7}},
  \bibinfo{pages}{457} (\bibinfo{year}{1992}).

\bibitem[{\citenamefont{Davis et~al.}(2007)}]{davis07}
\bibinfo{author}{\bibfnamefont{T.~M.} \bibnamefont{Davis}} \bibnamefont{et~al.}
  (\bibinfo{year}{2007}), \eprint{astro-ph/0701510}.

\bibitem[{\citenamefont{Spergel et~al.}(2006)}]{WMAP06}
\bibinfo{author}{\bibfnamefont{D.~N.} \bibnamefont{Spergel}}
  \bibnamefont{et~al.} (\bibinfo{collaboration}{WMAP}) (\bibinfo{year}{2006}),
  \eprint{astro-ph/0603449}.

\end{thebibliography}

\end{document}